\documentclass[final]{jfm}
\usepackage{amsfonts,amsbsy,amsgen,amssymb,natbib}
\usepackage[dvips]{graphics}
\usepackage{subfigure}
\usepackage{psfrag}
\usepackage[utf8]{inputenc} 
\usepackage[T1]{fontenc}
\usepackage{amsfonts,amsbsy,amsgen,amssymb}
\usepackage{amsmath}
\usepackage{epsfig}
\usepackage{color}

\renewcommand{\vec}{\mathbf}
\newcommand{\tens}[1]{\boldsymbol{\mathsf{#1}}}
\title[Matched-asymptotic expansions for determining the force acting on a particle]{Note on the method of matched-asymptotic expansions for determining the force acting on a particle}
\author[F. Candelier, R. Mehaddi and O. Vauquelin]
{\textbf{Fabien Candelier$\dagger$, Rabah Mehaddi and Olivier Vauquelin}}
\affiliation{
Aix-Marseille Universit\'e,  Laboratoire IUSTI, UMR CNRS 7343\\
5 rue Enrico Fermi, 13 453 Marseille Cedex 13, France}

\begin{document}
\maketitle
\begin{abstract}
 This paper is an addendum to the article by Candelier, Mehaddi \& Vauquelin (2013) where the motion of a particle in a stratified fluid is investigated theoretically, at small Reynolds and Péclet numbers.  We review briefly the method of matched asymptotic expansions which is generally used in order to determine the force acting on a particle embedded in a given flow, in order to account for small, but finite,  inertia effects.  As part of this method, we present an alternative matching procedure, which is based on a series expansion of the far-field solution of the problem, performed in the sense of generalized functions. The way to perform such a series is presented succinctly and a simple example is provided. 
\end{abstract}

\section{Introduction}

The prediction of the trajectories followed by small spherical inclusions embedded in a given flow has been, and remains, a topic of intense interest. This problem indeed has  obvious applications in many fields of physics, as well as in engineering science. In order to determine accurately these trajectories, a detailed understanding of the force acting on the particles is required. 

In the case  where the non-linear convective terms, involved in the Navier-Stokes equations written in a frame of reference moving with the particle, can be totally neglected, the force acting on an isolated solid sphere immersed in a Newtonian  fluid  is provided by the well-known Basset-Boussinesq-Oseen (BBO) equation (Boussinesq 1885; Basset 1888). The various terms appearing in this equation are the drag, the added-mass force and the history force. An equivalent equation for fluid inclusions immersed in an unsteady uniform flow (in the absence of inertia effects) has been obtained by Gorodtsov (1975) (see also Yang \& Leal 1991; Galindo \& Gerbeth 1993).  

However, the creeping-flow assumption is no longer valid in many situations. For example, when an inclusion is released into a quiescent fluid, the unsteadiness of the velocity perturbation eventually vanishes while convective terms are no longer negligible far from the inclusion (Oseen's problem). The determination of the particle-induced flow therefore requires us to solve a steady equation, in which linearised convective terms are involved (see Proudman \& Pearson 1957). As a result, it is found that the drag on the sphere is enhanced.  Another striking example is provided by the lift force which appears when a particle is embedded in a pure shear flow, owing to convective inertia effects (Saffman 1965, 1968; McLaughlin 1991; Asmolov \& McLaughlin 1999; Candelier \& Souhar 2007), as well as a shear-induced drag correction (see Harper \& Chang 1968; Miyazaki, Bedeaux \& Bonet Avalos 1995). Similarly, when particles are immersed in a solid-body rotation flow, convective inertia effects induce  lift forces,  as well as drag corrections in both the axial and the radial directions (Childress 1964; Herron, Davis \& Bretherton 1975; Gotoh 1990; Candelier, Angilella \& Souhar 2005; Candelier \& Angilella 2006; Candelier 2008). 

In order to determine the influence of inertia terms on the force acting on a particle, the method of matched asymptotic 
expansions is generally used, and more specifically, that devised  by Childress (1964) (see also Saffman 1965) and generalized to fluid inclusion by Legendre \& Magnaudet (1997). In this paper, the classical method is first presented in a general way, and then an alternative matching procedure, which is based on a series expansion of the far-field solution of the problem, performed in the sense of generalized functions, is proposed. 

\section{Description of the classical method}

Formally, in the problems mentioned in the introduction, the non-dimensional perturbed fluid motion equations, in which, in particular, lengths are scaled by the radius of the particle, can be written as follows
\begin{eqnarray}
\boldsymbol{\nabla} \cdot \vec{w} &=& 0 \:,\label{div}\\
- \boldsymbol{\nabla} p +\boldsymbol{\triangle} \vec{w} &=& \vec{q}_\epsilon\:, \label{eq_gene2}\\
\vec{w} = \vec{u}_r \mbox{ on }   r=1\:, \quad &\mbox{and}& \quad \vec{w} \to 0 \mbox{ as } r \to \infty\:,
\label{eq_gene3}
\end{eqnarray}
where $\vec{u}_r$ is the relative velocity of the particle (i.e. $\vec{u} - \vec{v}$, where $\vec{u}$ is the velocity of the particle,  $\vec{v}$, that of the fluid, and where the rotation of the particle has been neglected).   In this kind of problem, the creeping flow equations are perturbed by a term $\vec{q}_\epsilon$, which is such that
\begin{equation}
\lim_{\epsilon \to 0} \vec{q}_\epsilon  \to 0\:,
\label{vanish}
\end{equation}
and whose analytical expression naturally depends on the case considered.

In  a region close to the particle, i.e. characterized by $r \sim 1$ , and which is usually called the inner zone,  the solution of  equations (\ref{div}) to (\ref{eq_gene3}) is expanded formally as 
\begin{equation}
\vec{w}= \vec{w}_0 + \epsilon \vec{w}_1 + O(\epsilon^2) \quad \mbox{and} \quad p = p_0 + \epsilon \:p_1 + O(\epsilon^2)
\label{expansion_inner}
\end{equation}
where the zeroth-order velocity and pressure satisfy the creeping flow equations (i.e. $\vec{q}_\epsilon = 0$). 
As shown by Saffman (1965),  the inner problem is generally not regular since the boundary condition at infinity cannot be satisfied by the term  $\vec{w}_1$ which is found to be of order $O(r)$  for large $r$ in the great majority of cases.  As a consequence,  the first-order correction terms have to be matched to an outer solution, i.e. a solution which valid far from particle (i.e. $r \gg 1$). 

The far-field solution   is obtained by considering that in this region, the inclusion is seen by the fluid as a punctual force, modelled by Dirac-source term whose strength corresponds to that of a  Stokes drag (with a minus sign), and which leads us to
\begin{equation}
- \boldsymbol{\nabla} p + \boldsymbol{\triangle} \vec{w} + 6\:\pi\: \vec{u}_r \delta(\vec{x}) = \vec{q}_\epsilon \:.
\label{eq_far_field}
\end{equation}
In terms of  stretched coordinates $\tilde{\vec{x}} = \epsilon \vec{x}$, and after noticing that $\delta(\vec{x}) = \epsilon^3 \delta(\tilde{\vec{x}})$, equation (\ref{eq_far_field}) can be re-written as follows 
\begin{equation}
- \tilde{\boldsymbol{\nabla}} p' + \tilde{\boldsymbol{\triangle} } \vec{w}'  + 6\:\pi\: \vec{u}_r  \epsilon\:\delta(\tilde{\vec{x}})= \vec{q}'_{1} \:,
\label{eq_far_field2}
\end{equation}
where the fluid velocity, the pressure and the perturbation term, are now are denoted with a prime, in order to distinguish them from the variables written in normal coordinates. 

Let us denote by $\epsilon \: \vec{w}_{\mbox{\scriptsize out}}'$  the (normalized) solution of  equation (\ref{eq_far_field2}). For later convenience, let us also introduce  $\epsilon \: \vec{w}_{St}'$, the well-known Stokeslet solution (here written in terms of stretched-coordinates) which simply corresponds to the solution of (\ref{eq_far_field2}) in the particular case $\vec{q}'_{1}=0$. 

The last step of the method consists in matching the inner and the outer solution in a region where both the solutions are supposed to be valid. Such a  region is actually defined by $r \sim 1/\epsilon$, where we should  have
\begin{equation}
\vec{w}_0 + \epsilon \vec{w}_1  \sim  \epsilon \vec{w}_{\mbox{\scriptsize out}}'\:.
\label{raccordement}
\end{equation}
In view of the fact that in the matching region $r \gg 1$,  the velocity $\vec{w}_0$ naturally tends to the Stokeslet solution, i.e. $\vec{w}_0 \to \epsilon \vec{w}'_{St}$, it can be inferred, after simplifiying the parameter $\epsilon$ that 
$$
\vec{w}_1(\vec{x})  \sim   \vec{w}_{\mbox{\scriptsize out}}'(\tilde{\vec{x}}) - \vec{w}'_{St}(\tilde{\vec{x}})\:.
$$
After rewriting $\vec{w}_1(\vec{x})$ and $\vec{w}'(\tilde{\vec{x}}) - \vec{w}'_{St}(\tilde{\vec{x}})$ in terms of an intermediary variable  (see for instance Hinch 1991)
$$
\boldsymbol{\eta} = \frac{\vec{\tilde{x}}}{\epsilon^{1-\alpha}} = \epsilon^{\alpha} \vec{x}\:, \quad \mbox{where} \quad 0 < \alpha < 1\:,
$$
and by  taking the limit when $\epsilon \to 0$ for a fixed value of $\boldsymbol{\eta}$, we are led to the following matching condition
\begin{equation}
\lim_{|\vec{x}| \to \infty} \vec{w}_1(\vec{x})  = \lim_{|\tilde{\vec{x}|}\to 0}  (\vec{w}_{\mbox{\scriptsize out}}'(\tilde{\vec{x}}) - \vec{w}'_{St}(\tilde{\vec{x}}))\:.
\label{mathching}
\end{equation}
 
Note that in general, the  solution of (\ref{eq_far_field2}) is obtained by using Fourier transforms, in order to deal with the Dirac-source term involved in it. Thus, by defining the Fourier transform as follows
\begin{equation}
\vec{\hat{f}}(\vec{k}) = \int_{\mathbb{R}^3} \vec{f}(\tilde{\vec{x}}) \exp(-i\:\vec{k}\cdot \tilde{\vec{x}} ) \mbox{d} \tilde{\vec{x}} \:,
\label{def_Fourier}
\end{equation}
where $i$ is the imaginary unit (i.e. $i^2=-1$), and  the inverse Fourier transform by
$$
\vec{f}(\vec{x},\:t) = \frac{1}{(2\pi)^3} \int_{\mathbb{R}^3}\vec{\hat{f}}(\vec{k}) \exp(i\:\vec{k}\cdot \tilde{\vec{x}} )\mbox{d} \vec{k} \:,
$$
 equation (\ref{mathching}) generally reads as
\begin{equation}
\lim_{|\vec{x}| \to \infty} \vec{w}_1(\vec{x})  =  \frac{1}{(2\pi)^3}  \int_{\mathbb{R}^3} ( \hat{\vec{w}}_{\mbox{\scriptsize out}}'(\vec{k}) - \hat{\vec{w}}'_{St}(\vec{k}))\:\mbox{d} \vec{k} \:.
\label{matching2}
\end{equation}
Physically, (\ref{matching2}) means  that in the matching region,  the perturbation term $\vec{w}_1$ matches a uniform stream of velocity. In practice, this uniform flow is linked to the relative velocity of the particle by a linear relation of the form
$$
 \frac{1}{(2\pi)^3}  \int_{\mathbb{R}^3} ( \hat{\vec{w}}_{\mbox{\scriptsize out}}'(\vec{k}) - \hat{\vec{w}}'_{St}(\vec{k}))\:\mbox{d} \vec{k}= -\tens{M} \cdot \vec{u}_r
$$
where $\tens{M}$ defines a mobility-like  tensor.
 In return,  this outer uniform flow exerts on the particle, an additional Stokes drag so that the force acting on the particle finally reads as
$$
\vec{f}_1 = -6 \pi (\tens{I}  + \epsilon \tens{M}) \cdot \vec{u}_r\:,
$$
where $\tens{I}$ is the identity tensor. 

As discussed in the introduction of the paper, the method of matched asymptotic expansions has been applied successfully in many physical situations. However, in some cases, the integral  involved in the matching procedure  (\ref{matching2}) may be difficult to solve, owing to the complexity of the analytical expression of its integrand. In what follows, an alternative matching procedure is proposed, which can help to perform the matching between the inner and the outer expansions in the cases where the classical procedure fails.

\section{Series expansions of generalized functions}

The alternative matching procedure is based on series expansions of generalized functions. Let us then first recall basic definitions and fundamental results concerning distributions. 

Suppose that $f(\vec{x}):\:\mathbb{R}^3 \to \mathbb{R}$ is a locally integrable function, and let $\phi(\vec{x}):\:\mathbb{R}^3 \to \mathbb{R}$ be a test function in the Schwartz space  $\mathcal{S}(\mathbb{R}^3)$ (function space of all infinitely differentiable functions that are rapidly decreasing at infinity). The tempered distribution $\mathcal{T}_f$ which corresponds to the function $f$ is defined by
$$
\left< \mathcal{T}_f \:,\phi\right>  = \int_{\mathbb{R}^3} f(\vec{x}) \phi(\vec{x}) \mbox{d} \vec{x} \:.
$$
According to this definition, several other mathematical tools can be defined, as for instance, the partial derivative of a distribution with respect to a spatial coordinates, say $x_i$,   
\begin{equation}
\left< \frac{\partial \mathcal{T}_f}{\partial x_i} \:,\phi\right>  \triangleq  - \left< \mathcal{T}_f\:,\: \frac{\partial \phi}{\partial x_i}\right> \:.
\end{equation} 
Also, the Fourier transform of a distribution, that we shall denote by $\mathcal{F}(\mathcal{T}) = \hat{\mathcal{T}}_{f}$ can be defined by
\begin{equation}
\left< \hat{\mathcal{T}}_{f}\:,\:\phi \right>  \triangleq \left< \mathcal{T}_f\:,\:\hat{\phi} \right> \:.
\end{equation}
Note that the symbol $ \triangleq$ used in these two last equations stands for  'equal by definition'.

\vspace{11pt} 

Now in the case where the function $f(\vec{x})$  is integrable over the whole space, i.e. 
\begin{equation}
\int_{\mathbb{R}^3} f(\vec{x})  \mbox{d} \vec{x}  = C\:, \quad \mbox{where $C$ is a constant} 
\label{int_f}
\end{equation}
a fundamental result is that, in the sense of generalized function, 
\begin{equation}
\lim_{\epsilon \to 0} \frac{1}{\epsilon^3} \:f\left(\frac{\vec{x}}{\epsilon}\right)  \to  C\:\delta(\vec{x})\:.
\label{prop1}
\end{equation}
The demonstration of this result is rather simple and it is instructive to examine it (see for instance Boccara 1997). Briefly, to demonstrate (\ref{prop1}), stretched coordinates $\tilde{\vec{x}} = \epsilon \vec{x}$ are generally introduced, which allows the function
$$
g_\epsilon(\tilde{\vec{x}} ) =   \frac{1}{\epsilon^3} \:f\left(\frac{\vec{\tilde{x}}}{\epsilon}\right)\:,
$$
to be defined. 
By  a change of variable in  (\ref{int_f}), one can readily verify that, 
\begin{equation}
\int_{\mathbb{R}^3} g_{\epsilon} \:\mbox{d} \tilde{\vec{x}}  = C\:,
\end{equation}
so that the effect of  the distribution $\mathcal{T}_{g_\epsilon}$  on a test function $\phi$ can be arbitrarily  re-written as
$$
\left< \mathcal{T}_{g_\epsilon} \:,\phi\right>   = \int_{\mathbb{R}^3}  g_\epsilon(\tilde{\vec{x}}) \Big(\phi(\tilde{\vec{x}})- \phi(0) \Big) \mbox{d} \tilde{\vec{x}} + C \phi(0) \:.
$$
Taking the limit when $\epsilon \to 0$, and re-introducing unstretched coordinates therefore yields
$$
\lim_{\epsilon \to 0} \left< T_{g_\epsilon} \:,\phi\right>  = \lim_{\epsilon \to 0}  \int_{\mathbb{R}^3}  f(\vec{x}) \Big(\phi(\tilde{\vec{x}})- \phi(0) \Big) \mbox{d} \vec{x} + C \phi(0) \:.
$$
For a fixed value of $\vec{x}$, and because $f$ is locally integrable,  the integral term vanishes when $\epsilon \to 0$, so that
$$
\lim_{\epsilon \to 0} \left< T_{g_\epsilon} \:,\phi\right>  = C  \phi(0)\quad \Longleftrightarrow \quad \lim_{\epsilon \to 0}  T_{g_\epsilon} = C \delta\:.
$$

To take the analyse one step further, let us now consider a generalized function, say  $\mathcal{T}$, which  is defined by 
\begin{equation}
\mathcal{T} = \lim_{\epsilon\to 0} \frac{1}{\epsilon^{3+k}} \:f\left(\frac{\vec{x}}{\epsilon}\right)  \:,
\label{def_T}
\end{equation}
where $k$ is an arbitrary integer. Let us further consider three other integers, $\ell$, $m$ and $n$, which  are  such that $\ell+m+n=k$.
According to  (\ref{prop1}), it can be inferred that, for any combination of  $\ell$, $m$ and $n$, we should have
\begin{equation}
x_1^\ell \:x_2^m \:x_3^n \mathcal{T}    = C \: \delta(\vec{x}) \:,   \quad \mbox{where} \quad C = \int_{\mathbb{R}^3}  x_1^\ell\: x_2^m\: \:x_3^n\:f(\vec{x})  \mbox{d} \vec{x} \:.
\label{eq_distribution}
\end{equation}
 In terms of generalized function, solving equation (\ref{eq_distribution}) leads us to 
\begin{equation}
\mathcal{T}    = C \frac{(-1)^k}{\ell ! m! n!} \frac{\partial^k \delta}{\partial x_1^\ell \partial x_2^n \partial x_3^m}\:. 
\end{equation}
Note that any derivative of the delta distribution of lower order than $k$  is also a solution of (\ref{eq_distribution}), however such terms are not compatible with (\ref{def_T}) so that they should be zero.

By taking into account every possible combination, we are finally led to the following results:
\begin{itemize}
\item in the case $k=1$ 
$$
\lim_{\epsilon\to 0} \frac{1}{\epsilon^{4}} \:f\left(\frac{\vec{x}}{\epsilon}\right)  = \sum_{i=1}^3 C_i \frac{\partial \delta}{\partial x_i} \quad \mbox{with} \quad  C_i = \int_{\mathbb{R}^3}  x_i\:f(\vec{x})  \mbox{d} \vec{x} \:,
$$
\item in the case $k=2$
$$
\lim_{\epsilon\to 0} \frac{1}{\epsilon^{4}} \:f\left(\frac{\vec{x}}{\epsilon}\right)  = \sum_{i=1}^3 \sum_{j\geq i}^{3} C_{ij} \frac{\partial^2 \delta}{\partial x_i \partial x_j} \:,
$$
$$
 \mbox{with } C_{ij} = \int_{\mathbb{R}^3}  x_i x_j\:f(\vec{x})  \mbox{d} \vec{x}  \mbox{ if $i \neq j$, and else } 
 C_{ii} = \frac{1}{2} \int_{\mathbb{R}^3}  x_i^2 \:f(\vec{x})  \mbox{d} \vec{x}  \:,
$$
\item and so on
$$
\lim_{\epsilon \to 0} \frac{1}{\epsilon^{3+k}} \:f\left(\frac{\vec{x}}{\epsilon}\right)  \to  \underbrace{\sum_{i_1=1}^{3}\sum_{i_2\geq i_1}^{3} \ldots 
\sum_{i_n\geq i_{n-1}}^{3}}_{k \:\mbox{\scriptsize times}} C_{i_1i_2 \ldots i_n}\:\frac{\partial^k \delta(\vec{x})}{\partial x_{i_1}\partial x_{i_2} \ldots \partial x_{i_n}} \:,
$$
$$
\mbox{with }
C_{i_1i_2 \ldots i_n}= \frac{(-1)^k}{\ell ! m! n!} 
\int_{\mathbb{R}^3}  x_1^\ell\: x_2^m \:x_3^n\:f(\vec{x})  \mbox{d} \vec{x} \:,
$$
where $\ell$, $m$ and $n$  are determined by the number of occurrences, respectively,  of the indices $1$, $2$ and $3$ in the sequence $i_1i_2 \ldots i_n$. 
\end{itemize}

\subsection{A simple example}
In order to  illustrate how  such results can be used to approximate a perturbed form of the Green function of a differential equation, let us first consider  a very simple example based on a steady Schr{\"o}dinger-like equation of the form :
$$
\triangle f - \epsilon^2\:f= \delta \:.
$$
To determine the Green function of this equation, the (spatial) Fourier transform defined in (\ref{def_Fourier}) can be used, which yields
\begin{equation}
\hat{f} = -\frac{1}{k^2+\epsilon^2} \:,
\label{pot_pertube}
\end{equation}
and then,  calculating the  inverse Fourier transform of $\hat{f}$ leads us to
$$
\mathcal{F}^{-1}(\hat{f}) = f(\vec{x}) = -\frac{\exp(-\epsilon \:r) }{4\pi r}  \quad \mbox{where} \quad r=|\vec{x}|\:.
$$
This solution can be expanded, with respect to $\epsilon$, as  follows
\begin{equation}
-\frac{\exp(-\epsilon \:r) }{4\pi r} = -\frac{1}{4\pi\:r} + \frac{\epsilon}{4\pi} - \frac{r \epsilon^2}{8\pi} + \frac{r^2 \epsilon ^3}{24 \pi} + O(\epsilon^4)\:.
\label{expansion}
\end{equation}
Obviously, such a series could not have been  retrieved directly  from a naive expansion of (\ref{pot_pertube}) (i.e.  performed in the sense of classical function) since only powers of 2 would be involved in it. In contrast, (\ref{expansion}) can be retrieved  if the series is performed in the sense of generalized function. Similarly as for classical functions, such a series reads as  
$$
 -\frac{1}{k^2+\epsilon^2} = \hat{\mathcal{T}}_0 + \epsilon \:  \hat{\mathcal{T}}_1 + \epsilon^2 \:  \hat{\mathcal{T}}_2 + \ldots  \epsilon^n \:  \hat{\mathcal{T}}_n
 + O(\epsilon^{n+1})
 $$
 where 
\begin{equation}
 \mathcal{T}_n  = \frac{1}{n! }\lim_{\epsilon \to 0} \frac{\mbox{d}^n }{\mbox{d} \epsilon^n}  \left( -\frac{1}{k^2+\epsilon^2}\right) \:.
\label{eq_det_tn}
 \end{equation}
In our simple example, by calculating the first term, we are led to
 $$
  \hat{\mathcal{T}}_0 = \lim_{\epsilon \to 0}   -\frac{1}{k^2+\epsilon^2} = -\frac{1}{k^2} 
\quad  \mbox{and} \quad  
 \mathcal{F}^{-1}(\hat{\mathcal{T}}_0) = -\frac{1}{4\pi\:r}\:.
$$
For the second term, 
$$
 \hat{\mathcal{T}}_1 = \lim_{\epsilon \to 0} \frac{\mbox{d}}{\mbox{d} \epsilon} \left( -\frac{1}{k^2+\epsilon^2} \right) = \lim_{\epsilon \to 0} \frac{1}{\epsilon^3} \frac{2}{((k/\epsilon)^2 + 1)^2}\:.
$$
According to the fact that
$$
\int_{\mathbb{R}^3}  \frac{2}{((k^2 + 1)^2} \mbox{d}\vec{k} = 2\pi^2\:,
$$
this yields
$$
\hat{\mathcal{T}}_1 = 2 \pi^2 \delta(\vec{k}) \quad \mbox{and} \quad  \mathcal{F}^{-1}(\hat{\mathcal{T}}_1) = \frac{1}{4\pi}\:,
$$
since $\mathcal{F}^{-1}(\delta) = 1/(8\pi^3)$. 
By pursuing the expansion in a similar way, we are finally led to 
$$
\hat{\mathcal{T}}_2 =  \frac{1}{k^4} \quad \mbox{and} \quad \mathcal{F}^{-1}(\hat{\mathcal{T}}_2) = -  \frac{r}{8\pi}\:, 
$$
$$
\hat{\mathcal{T}}_3 = - \frac{\pi^2}{3} \Delta_k \delta(\vec{k}) \quad \mbox{and} \quad \mathcal{F}^{-1}(\hat{\mathcal{T}}_3) = \frac{r^2}{24\pi}\:,
\quad \mbox{etc} \ldots
$$
so that  (\ref{expansion}) is indeed recovered. 

%
%

\section{The alternative matching procedure}

According to the results presented in the previous section, an alternative matching procedure  can now be proposed. Indeed, by 
considering the fact that  the parameter $\epsilon$ is small compared to unity, the Fourier transform of the solution of the (unstretched) outer equation (\ref{eq_far_field}) can be expanded in terms of generalized functions 
$$
\vec{\hat{w}}  = \hat{\mathcal{T}}_0 + \epsilon  \hat{\mathcal{T}}_1+ \epsilon^2 \:  \hat{\mathcal{T}}_2  + \ldots +  \epsilon^n \:  \hat{\mathcal{T}}_n 
$$
where similarly as in the previous case,  the generalized functions $\mathcal{T}_n$ are determined by 
\begin{equation}
 \mathcal{T}_n  = \frac{1}{n! }\lim_{\epsilon \to 0} \frac{\mbox{d}^n\vec{\hat{w}}   }{\mbox{d} \epsilon^n} \:.
\label{eq_det_tn}
 \end{equation}
%
 According to (\ref{vanish}), it is readily found that $\hat{\mathcal{T}}_0$  simply corresponds to the Fourier transform of a Stokeslet, so that, in the matching zone, its inverse (spatial) Fourier transform naturally matches  the leading order term $\vec{w}_0$ of the inner expansion  (\ref{expansion_inner}). 

In our problems, the second term is always found to be of the form
$$
\hat{\mathcal{T}}_1  = \lim_{\epsilon\to 0} \frac{1}{\epsilon^3} \:\vec{f}\left(\frac{\vec{k}}{\epsilon}\right)\:,
$$
which implies that 
$$
\hat{\mathcal{T}}_1  = \vec{C} \:\delta(\vec{k})  \quad \mbox{where} \quad   \vec{C}  = \int_{\mathbb{R}^3} \frac{\mbox{d} \vec{\hat{w}}  }{\mbox{d}\epsilon} \Big|_{\epsilon=1} \: \mbox{d} \vec{k}\:.
$$
By noticing that the inverse spatial Fourier transform of $\delta(\vec{k})$  is given by $1/(2\pi)^3$, 
and similarly as in the classical method,  is observed that in the matching zone $r\sim 1/\epsilon$,  the perturbation term $\vec{w}_1$ of the inner expansion (\ref{expansion_inner}) matches a uniform velocity stream given by $\vec{C}/(2\pi)^3$, and we are led to the same conclusions as for the classical matching procedure.

\section{Concluding remarks}

It is worth mentioning that formally 
$$
\frac{1}{8\pi^3}\int_{\mathbb{R}^3} \frac{\mbox{d} \vec{\hat{w}}  }{\mbox{d}\epsilon} \Big|_{\epsilon=1} \: \mbox{d} \vec{k} = 
 \int_{\mathbb{R}^3} ( \hat{\vec{w}}_{\mbox{\scriptsize out}}'(\vec{k}) - \hat{\vec{w}}'_{St}(\vec{k}))\:\mbox{d} \vec{k}
 $$
which means that the two matching procedures obviously provide us with the same result.  In some  ways, the difference between the two approaches can be viewed as an inversion between taking the limit when $\epsilon \to 0$, and then performing the integration, or conversely, performing first the integration,  and then taking the limit. Also, let us mention that the alternative matching procedure proposed here has been tested in many configurations where the classical method applies well as, in particular, the problem considered by Oseen (Proudman \& Person 1957) or that considered by  Herron {\it et al.} (1975).

Let us finally mention that this alternative method has been specifically developed to allows us to determine the drag correction induced by the flow perturbation on a particle in the problem recently addressed by Ardekani \& Stoker (2010). These authors have investigated  the flow produced by a point force, intended to represent a settling 
particle, in a stratified fluid, at small Reynolds and Péclet numbers. In particular, in this study, the  creeping flow solution is perturbed by buoyancy effects, and therefore, it can be cast into the formalism described in \S 2, except that the integral involved in the classical matching procedure  (\ref{matching2}) cannot be solved analytically,  owing to the complexity of the analytical expression of its integrand. 
These results, which have also been generalized to the unsteady case, have been the subject  of a companion paper by Candelier, Mehaddi \& Vauquelin 2013.



\begin{thebibliography}{natbib}

\bibitem[Ardekani \& Stoker(2010)]{Ardekani10}
{Ardekani, A. M. \& Stocker, R.} 2010 {Stratlets: Low Reynolds Number Point-Force Solutions in a Stratified Fluid.} \textit{Phys. Rev. Letter}
\textbf{105}, 084502

\bibitem[Asmolov \& McLaughlin(1999)]{Asmolov99}
{Asmolov, E. \& McLaughlin, J.B.} 1999 {The inertial lift on a
oscillating sphere in a linear shear flow.} \textit{Int. Journal
of Multiphase Flow} \textbf{183}, 199--218.


\bibitem[Basset (1888)]{Basset88}
{Basset, A. B.} 1888 {Treatise on hydrodynamics.} \textit{Deighton
Bell, London.} \textbf{2}, 285--297.

\bibitem[Boccara(1997)]{Boccara97}
{Boccara, N.} 1997 {Distributions.}
\textit{Ellipses,} Paris. pp 53--54. 

\bibitem[Boussinesq (1885)]{Boussinesq85}
{Boussinesq, J.} 1885 {Sur la r\'esistance qu'oppose un fluide
ind\'efini au repos sans pesanteur au mouvement vari\'e d'une
sph\`ere solide qu'il mouille sur toute sa surface quand les
vitesses restent bien continues et assez faibles pour que leurs
carr\'es et produits soient n\'egligeables.} \textit{C. R. Acad.
Sci. Paris} \textbf{100}, 935--937.

\bibitem[Candelier, Mehaddi \& Vauquelin(2013)]{Candelier13}
{Candelier, F.,  Mehaddi, R. \& Vauquelin, O.} 2013 {The history force on a small particle in a linearly stratified fluid} \textit{Under consideration for publication in J. Fluid Mech.}  






\bibitem[Candelier, Angilella \& Souhar(2005)]{Candelier05}
{Candelier, F., Angilella, J.-R. \& Souhar, M.} 2005 {On the
effect of inertia and history forces on the slow motion of a
spherical solid or gaseous inclusion in a solid-body rotation
flow.} \textit{J. Fluid. Mech.} \textbf{545}, 113--139.

\bibitem[Candelier \& Angilella(2006)]{Candelier06}
{Candelier, F. \& Angilella, J. R.} 2006 {Analytical investigation
of the combined effect of fluid inertia and unsteadiness on low-Re
particle centrifugation.} \textit{Phys. Rev. E} \textbf{73},
047301.

\bibitem[Candelier \& Souhar(2007)]{Candelier07}
{Candelier, F. \& Souhar, M.} 2007 {Time-dependent lift force
acting on a particle moving arbitrarily in a pure shear flow, at
small Reynolds number} \textit{Phys. Rev. E} \textbf{76}, 067301.

\bibitem[Candelier(2008)]{Candelier08}
{Candelier, F.} 2008 {Time-dependent force
acting on a particle moving arbitrarily in a rotating flow, at small Reynolds and Taylor numbers.} \textit{J. Fluid. Mech.} \textbf{608}, 319--336.


\bibitem[Childress(1964)]{Childress64}
{Childress, S.} 1964 {The slow motion of a sphere in a rotating,
viscous fluid.} \textit{J. Fluid Mech.} \textbf{20}, 305--314.


\bibitem[Galindo \& Gerbeth (1993)]{Galindo93}
{Galindo, V. \& Gerbeth, G. } 1993 {A note on the force on an
accelerating spherical drop at low Reynolds number.} \textit{Phys.
Fluids} \textbf{5(12)}, 3290--3292.

\bibitem[Gorodtsov (1975)]{Gorodtsov75}
{Gorodtsov, V. A.} 1975 {Slow motions of a liquid drop in a
viscous liquid.} \textit{Zhurnal Prikladnoi Mekhaniki i
Tekhnicheskoi Fiziki} \textbf{6(2)}, 32--37. (translated in : J.
Appl. Mech. Tech. Phys. {\bf(16)}, 865--868)

\bibitem[Gotoh (1990)]{Gotoh90}
{Gotoh, T.} 1990 {Brownian motion in a rotating flow} \textit{J.
Stat. Phys.} \textbf{59}, 371--402.

\bibitem[Harper \& Chang(1968)]{Harper68}
{Harper, E. Y. \& Chang, I-Dee.} 1968 {Maximum dissipation resulting from lift in a slow viscous shear flow.} \textit{J. Fluid Mech.}
\textbf{33 (2)}, 209--225.

\bibitem[Herron {\it et al.} (1975)]{Herron75}
{Herron, I. H., Davis, S. \& Bretherton F. P.} 1975 {On the
sedimentation of a sphere in a centrifuge.} \textit{J. Fluid
Mech.} \textbf{68}, part. 2, 209--234.


\bibitem[Hinch(1991)]{Happel91}
{Hinch, E. J.} 1991 {Perturbation methods.}
\textit{Cambridge university press}.

\bibitem[Legendre \& Magnaudet(1997)]{Legendre97}
{Legendre, D. \& Magnaudet, J.} 1997  {A note on the lift force on a spherical bubble or drop in a low-Reynolds shear flow} \textit{Phys. Fluids} \textbf{9}(11),
3572--3574.



\bibitem[McLaughlin (1991)]{McLaughlin91} {McLaughlin, J. B.}
1991 {Inertial migration of a small sphere in linear shear flows.}
\textit{J. Fluid Mech.} \textbf{224}, 261--274.

\bibitem[Miyazaki {\it et al.}(1995)]{Miyazaki95b}
{Miyazaki, K., Bedeaux, D. \& Bonnet Avalos, J.} 1995 {Drag on a
sphere in a slow shear flow.} \textit{J. Fluid Mech.}
\textbf{296}, 373--390 (1995).

\bibitem[Proudman \& Pearson  (1957)]{Proudman57}
{Proudman, I. \& Pearson, J. R. A. } 1957 {Expansions at small
Reynolds numbers for the flow past a sphere and circular
cylinder.} \textit{J. Fluid Mech.} \textbf{2}, 237--262.

\bibitem[Saffman(1965)]{Saffman65}
{Saffman, P.G.} 1965 {The lift on a small sphere in a viscous flow.} \textit{J. Fluid Mech.} \textbf{22}, part 2, 385--400
\textit{and corrigendum (1968)}.


\bibitem[Yang \& Leal (1991)]{Yang91} {Yang, S. \& Leal, L. G.} 1991 {A note on memory integral contributions
to the force on an accelerating spherical drop at low Reynolds
number.} \textit{Phys. Fluids A} \textbf{3}(7), 1822--1824.


\end{thebibliography}
\end{document}